\documentstyle[aps,psfig]{revtex}
\def\am{angular momentum\  }

\begin{document}
\author{V.I.Dimitrov, F. D\"onau and S. Frauendorf\\
Institute for Nuclear and Hadronic Physics,
Research Center Rossendorf, PB 51 01 19,  01314 Dresden, Germany\\
and \\ Department of Physics, University of Notre Dame, 
Notre Dame, IN 46556, USA}
\title{A hybrid version of the tilted axis cranking model
and its application to $^{128}$Ba }
\maketitle
 
\abstract{A hybrid version the deformed nuclear potential is suggested,
which combines a spherical Woods Saxon potential with a deformed 
Nilsson potential. It removes the problems of the conventional Nilsson 
potential in the mass 130 region. 
Based on the hybrid potential, 
tilted axis cranking calculations are carried out  for the
 magnetic dipole band  in  $^{128}$Ba.         }
\section{Introduction}

The transitional nuclei in the region $A=130-140$ show regular
 $\Delta I=1$ regular bands, characterized by large $
B(M1)/B(E2)$ ratios, the lack of signature splitting and relatively low
dynamical moments of inertia \cite{Ref1}.
These bands have an  intermediate character. They stand between the 
collective high-$K$ bands of well deformed nuclei, for which 
collective rotation is the dominant mechanism
of generating the \am, and the magnetic rotation of near spherical nuclei, 
for which few high-$j$ particles and
holes generate most of the \am by means of the   shears mechanism
(see, for example \cite{zphys,rmp}).    
The question of how changes magnetic  into collective rotation 
has not been studied yet.
So far, only the magnetic dipole band in  $^{128}$Ba has been investigated
\cite{Ref7} from this point of view. 
  Another intriguing question is
the possibility of a chiral character of rotation \cite{chiral}.
Their softness  with respect to 
triaxial deformations makes the nuclei in the $A=130$ region 
 particularly good candidates for identifying  this
 new symmetry type.

The Tilted Axis Cranking (TAC) model \cite{Ref4} has  turned out to be
an appropriate
theoretical tool for the description of the magnetic dipole bands.
 This model is
a natural generalization of the cranking model  \cite{Ref5} for
situations where the axis of rotation does not coincide with a principal
axis of the density distribution of the rotating nucleus, and thus the
signature is not a good quantum number. Since  introduced, TAC
has proven to be  a reliable  approximation for the energies and intraband
transitions in both normally and weakly deformed nuclei \cite{rmp}. However,
in the case  of the four quasiparticle magnetic dipole 
band in $^{128}$Ba, the TAC calculations \cite{Ref7}
predicted the wrong 
parity and a too early termination of
the band  and provided only fair description of the electro magnetic
transition data \cite{Ref7a,Ref7b,Ref8}.   $^{128}$Ba  is one of
the best studied nuclei in this mass range and the above mentioned
$\Delta I=1$ rotational band
is a good test case for the TAC model. 

The purpose of the present work is to try to identify the origin of the
discrepancies and remove them.
The version of TAC used in \cite{Ref7} was based on the 
Nilsson Hamiltonian.
The parameters of this well-tried potential
were  carefully adjusted for various mass
regions, where   
it was successfully used in  standard cranking calculations
\cite{NilsRag} as well as in TAC calculations \cite{rmp}. 
However, the $A=130$ region is known to be problematic for the model.
In fact no really satisfactory set of Nilsson parameters for
this region is available so far\cite{Ref9}. 
Thus we attribute  the discrepancies  of the TAC calculations
 for $^{128}$Ba \cite{Ref7}
with the later measurements \cite{Ref7a,Ref7b,Ref8} to the general problems 
of the Nilsson potential in this region.   

On the other hand, the Woods-Saxon  potential  works
very well around $A=130$ \cite{Ref10}.
Encouraged by this,   we adapt the Nilsson potential 
as close as possible to the Woods-Saxon one. We call this the hybrid
potential, which is the basis for  new TAC calculations for $^{128}$Ba.   
Instead of the parameterizing the single particle levels of the spherical 
modified oscillator in the standard way  by means of an $ls$- and an 
$l^2$-term, the hybrid model directly takes the energies of the spherical
Woods-Saxon potential. The deformed part of the hybrid potential is 
an anisotropic harmonic oscillator. This compromise keeps the simplicity of
the Nilsson potential, because coupling between the oscillator shell can
be approximately taken into account by means of stretched coordinates
\cite{NilsRag}, and it amounts to a minor modification of the existing TAC code.
On the other hand, it has turned out to be a quite good approximation 
of the realistic flat bottom potential as long as the deformation is
moderate. The hybrid was used to calculate the triaxial shapes of 
liquid sodium clusters \cite{clhyb}. The results agree very well
with  later calculations using the correct radial profile of the 
deformed part of the potential\cite{clks}.

\section{TAC implementation}

The TAC model is discussed in more detail in \cite{Ref4,zphys,rmp,tacdic}.
The brief 
presentation  in this paper focuses at the suggested  improvement 
of this approach.
The starting point  is the mean field Routhian\footnote{%
For simplicity, only one type of particles 
is spelled out. The extension to both types is obvious.} 
\begin{equation}
h^{\prime } = h_{sph} + V_{def}(\varepsilon_2,\gamma,\varepsilon_4)\\
-\Delta \left( P+P^{+}\right) -\lambda N \label{Eq.3} 
-\omega \left( \sin \vartheta \,j_{1}+\cos \vartheta \,j_{3}\right)
  \nonumber
\end{equation}
where $h_{sph}$  denotes the spherical part including the spin-orbit term 
and $ V_{def}(\varepsilon_2,\gamma,\varepsilon_4)$  the 
deformed part of  the  Nilsson 
single-particle Hamiltonian.
 (see, e.g. \cite{NilsRag}). 
The pairing field in eq.(1) is 
determined by the gap parameter $\Delta$ and 
the  monopole pairing operator  $P=\sum_{k}c_{k}c_{\overline{k}}$
while the chemical potential $\lambda N$ is needed to 
satisfy on average the particle number conservation.
The  two-dimensional cranking term 
$\omega \left( \sin \vartheta \,j_{1}+\cos \vartheta \,j_{3}\right) $ is
the  new element of the TAC, as compared to the standard cranking model,
which is recovered for $\vartheta$\,=\,0 or 90$^\circ$. 
The angle $\vartheta$ fixes the tilt of the cranking axis
 with respect to the intrinsic 3-axis
 within the principal
(1-3) plane  of the deformed potential.
By diagonalization in stretched coordinates,
neglecting $\Delta N=2$ shell mixing,  this 
TAC Routhian 
yields quasi-particle energies and quasi-particle states, from which the
many-body configuration  $
|\,\omega,\varepsilon_2,\gamma,\varepsilon_4,\vartheta> $ of interest is constructed.
 
The mean field is found for a given
frequency  $\omega$ and fixed configuration by minimizing
the total Routhian
\begin{equation}
E^\prime(\omega,\varepsilon_2,\gamma,\varepsilon_4,\vartheta)  =
 <\omega,\varepsilon_2,\gamma,\varepsilon_4,\vartheta\,|\,h^\prime\,|\,\omega,\varepsilon_2,\gamma,\varepsilon_4,\vartheta>
\end{equation}
with respect to the deformation parameters 
$(\varepsilon_2, \gamma,\varepsilon_4)$ and the tilt angle $\vartheta $. 
The value of $\Delta $ is kept fixed at two values:
 80\% of  the experimental odd-even mass
difference and zero. At the equilibrium angle 
$\vartheta=\vartheta_\circ$ (minimum)  the
cranking axis is parallel to the direction 
of the angular momentum vector $\vec J = (<j_1>,<j_3>)$. 
After the minimum is found
the various electro magnetic observables of interest 
are obtained by means of the 
following semiclassical expressions \cite{zphys,tacdic}
 \begin{eqnarray}
&<I-2I-2|{\cal M}_{-2}(E2)|II>=
<{\cal M}_{-2}(E2)>=&\\
&=\sqrt{{5\over4\pi}}\left(\frac{eZ}{A}\right)
\left[\sqrt{3\over8}<Q'_0>(\sin\,\vartheta)^2
+{1\over4}<Q'_2+Q'_{-2}>\left( 1+(\cos\vartheta)^2\right) \right],&\\
&<I-1I-1|{\cal M}_{-1}(E2)|II>=
<{\cal M}_{-1}(E2)>=&\\
&=\sqrt{{5\over4\pi}}\left(\frac{eZ}{A}\right)
\left[\sin\,\vartheta \cos\vartheta( \sqrt{3\over2}<Q'_0>
-{1\over2}<Q'_2+Q'_{-2}>)\right].&\\   
&<I-1I-1|{\cal M}_{-1}(M1)|II>=
<{\cal M}_{-1}(M1)>=&\\
&=\sqrt{\frac{3}{8\pi }}
\left[\mu_3\sin\,\vartheta -\mu_1\cos\,\vartheta \right],
\end{eqnarray}
where $<{\cal M}_{\nu}>$ is the expectation value 
of the transition operator  with the 
TAC configuration $|>$, the components of which refer to the lab system.   
The intrinsic quadrupole moments $Q'_\mu$ are calculated with respect to the 
principle axes (1, 2, 3). The same holds for the magnetic moments
\begin{equation}
\mu_1=\mu_N(J_{1,p}+(\eta 5.58-1)\,S_{1,p}-\eta 3.82\,S_{1,n}),~~
\mu_3=\mu_N(J_{3,p}+(\eta 5.58-1)\,S_{3,p}-
\eta 3.82 \,S_{3,n}),
\end{equation}
where the free nucleonic magnetic moments 
are attenuated by a factor of $\eta = 0.7$.
The reduced transition probabilities are
\begin{eqnarray*}
B(M1,\Delta I &=&1)=<{\cal M}_{-1}(M1)>^2\\
B(E2,\Delta I &=&2)=<{\cal M}_{-2}(E2)>^2
\end{eqnarray*}
The mixing ratio is
\begin{equation}
\delta=\frac{<{\cal M}_{-1}(E2)>}{<{\cal M}_{-1}(M1)>}.
\end{equation}

We  apply the Strutinsky renormalization procedure 
to calculate the  total Routhian 
\begin{equation}
E^{\prime }(\omega )=E_{LD}(\omega =0)-E_{smooth}+<\omega |\,h^{\prime
\,
}|\omega >  \label{Eq.5},
\end{equation}
where $E_{LD}=E_{LD}(\varepsilon_2,\gamma,\varepsilon_4,\varepsilon_4)$ 
means the liquid drop 
energy and $E_{smooth}$
is the 
the  smooth part of the mean-field energy calculated from the 
single-particle energies at $\omega =0$. 
This version of the TAC has turned out to be quite successful for well
deformed nuclei (see \cite{zphys,rmp,tacdic} and references therein).

The new element of the present paper consists in the 
hybrid potential, which approximates the well-established deformed 
Woods-Saxon potential, yet
preserving the existing convenient TAC environment. For this purpose   
 the spherical part $h_{sph}$ 
in Eq.(1) is replaced  the  spherical 
Woods-Saxon Hamiltonian for the nucleus of interest. 
In the present work the  universal Woods-Saxon parameters
 are used (see e.g. \cite{Ref13}).

Technically, the replacement is rather simple, because the existing TAC code
uses states of good $l,j,m$ as a basis. The spherical Nilsson
energies $e^{(nil)}_{N,l,j}$ are replaced by the spherical 
Woods - Saxon energies $e^{(ws)}_{N,l,j}$. It turns out to be unproblematic 
to associate the quantum numbers of the two different potentials. 
For a given combination $l,j$ 
the third quantum number $N$ is found by counting from the state with the
lowest energy. The fact that the spherical Woods-Saxon code
uses a harmonic oscillator basis permitted a check of the algorithm.
The major component of the Woods-Saxon wavefunction agrees with 
 the state found by our counting algorithm. 
In the high-lying part of the single-particle
spectrum (three shells above the valence shell or higher)
 there are occasional ambiguities in assigning the states.
Small errors of this kind are not expected to have any consequences
at moderate or small deformation. The states do not couple strongly
to the states near the Fermi surface. They contribute only to the 
smooth level density used in the Strutinsky renormalization,
which will not be affected by small shifts of the levels.

 Such a replacement of the spherical
single-particle energies is a common practice in large scale shell-model
configuration mixing and  similar calculations, where they
are often used as adjustable parameters \cite{Ref14}.  The effect of the
replacement is illustrated on Fig.1, which shows  the deformation 
dependence of the proton
single-particle levels of  $^{128}$Ba for the  Nilsson, 
the Woods-Saxon
 and the hybrid Hamiltonians. 
The close similarity of the levels of the Woods-Saxon and the hybrid models
is obvious. The hybrid has a  somewhat later and sharper
crossing between the positive parity levels,  which also show
stronger tendency to arrange into pairs of pseudo spin doublets.
These  treats are inherited from the Nilsson Hamiltonian,
which controls the change with deformation.
The main difference between the Nilsson Hamiltonian
and the other two is the  lower energy of the negative parity levels
 originating from 
$h_{11/2}$. 
It  seems to be the  reason for the discrepancies between
the previous TAC predictions and the experiment, as will be demonstrated in the next
section.

\section{The M1 band in $^{128}$Ba}
In the previous TAC calculations for the $\Delta I=1$ band in
$^{128}$Ba \cite{Ref7} the four quasi-particle configurations $[\pi
(h_{11/2})^{2}\nu (h_{11/2}(d_{5/2}g_{7/2}))]$
\footnote{Indicating
the major components, we denote  the 
mixed Nilsson state by $(d_{5/2}g_{7/2})$.} for the negative parity and $[\pi
(h_{11/2})^{2}\nu (h_{11/2})^{2}]$ for the positive parity 
were found to be the lowest ones in energy at the oblate 
deformation of $\varepsilon =0.26$, $\gamma =60^{o}$. 
This deformation was determined  at $\omega =0.2$ MeV by minimizing  
the total Routhian calculated from the Quadrupole-Quadrupole interaction.  
With the present
version of TAC we find a significantly  smaller  prolate
 equilibrium deformation of $\varepsilon
=0.205$ and $\gamma =0^{o}$. The lowest
four quasi-particle configuration now turns out to be $[\pi
(h_{11/2}(d_{5/2}g_{7/2}))\nu( h_{11/2}(d_{5/2}g_{7/2}))]$. Fig.2 shows the quasi proton
 and quasi neutron levels for various cranking
frequencies at tilt angle $\vartheta =90^{o}$ and for various angles at $\omega
=0.2$ MeV.
 The equilibrium value of $\vartheta $, which minimizes $E^{\prime}$,
 is found to be $\vartheta_\circ=52.5^{o}$ for $\omega=0.2$ MeV.   

As seen, the 
different position of the $h_{11/2}$ orbitals in the hybrid TAC 
has drastic consequences. The deformation changes from oblate
to prolate, resulting in a different configuration of the M1 band.
This is not surprising in a region where the energy difference 
between oblate and prolate shape is small. 

The calculations of the $\Delta I=1$ band in the
present work are built on this new configuration
 $[\pi(
h_{11/2}(d_{5/2}g_{7/2}))(\nu h_{11/2}(d_{5/2}g_{7/2}))]$, which is the lowest 
four-quasi particle TAC solution. In agreement with the experiment, it has
positive parity.
The excitation of four quasi-particles   significantly reduces
 the pairing gaps. In
order to better grasp the influence of this blocking on
characteristics of the band, we did two calculations: one with
 $\Delta _{\nu}=0.88$ MeV and $\Delta _{\pi }=1.04$ MeV 
corresponding to 80\% of the experimental 
even odd mass difference  and one with zero pairing.

Electro magnetic transition properties present a
 stringent test of the nuclear models.
The experimental information about lifetimes and  mixing ratios
of the magnetic dipole band
was accumulated in several experiments  \cite{Ref7a,Ref7b,Ref8}.
Fig.3 compares the experimental $B(M1)$ and $B(E2)$ values and mixing ratios 
 with the results of  TAC model obtained for the 
configuration  $[\pi( h_{11/2}(d_{5/2}g_{7/2}))\nu( h_{11/2}(d_{5/2}g_{7/2}))]$.
 All electro-magnetic 
characteristics of the band are 
well reproduced by both the paired and unpaired
calculations. Curiously, the quenching of  pairing
influences to some extent the $B(M1)$ values
but leaves almost unchanged the B(E2) values up to spin 18$\hbar$.
The TAC calculation seems to slightly overestimate the deformation.

In contrast
with the previous TAC calculation,  it is possible to follow the
band all the way up to spin 26$\hbar$. 
Fig.4 shows the measured and the calculated function $J(\omega)$ 
of the spin on the angular frequency.
 It is more sensitive to the changes 
of the pair correlations.
 While the unpaired calculation gives a nearly linear function with 
 the moment of inertia ${\cal J}^{(2)}=dJ/d\omega$ close to 
the measured one, the paired calculation exhibits a substantially lower
moment of inertia for low rotational frequencies and an
upbend at higher ones. The experiment is in between.
This seems to indicate that the pair field is weak in this nucleus
and a more refined treatment of pairing is needed. 
It is noted that in  the
calculations  the band extends down to
$\omega =0.1\, MeV$ and $J=11\hbar$. In experiment there is an irregularity
around $I=12,13\hbar$. It may be caused by mixing of the $13^+$ state with the
 another $13^+$ state,  which lies nearby in energy ($\Delta E=0.045\, MeV$)
and into which the $14^+$ also decays\cite{Ref8} .

In \cite{Ref7a,Ref7b,Ref8}, the $\Delta I=1$ band was  analyzed 
in terms of a pure high-K band using the familiar
expressions for the $B(M1)$ and $B(E2)$ values for the axial symmetric
rotor \cite{BM}. Adjusting three free parameters, 
the K-value, the intrinsic quadrupole moment $Q_\circ$, and
the gyromagnetic factor  $|g_K-g_R|$, a good fit of
 the electro magnetic 
decay data was obtained.  The quality  is 
practically the same as in our calculation without  parameters
 in Fig.4.  The TAC calculation contains much more physical information
as e.g.  the specific configuration on which the band is built
and  the band energies.
The  knowledge of the intrinsic state can be used 
to derive further structure information,  
e. g. the
geometrical coupling scheme  shown in Fig.5, 
which enables one to see how the total spin 
is formed from the quasiparticle orbitals and how it changes 
with the rotational frequency.   
Apparently, most of the angular momentum gain along the band is of 
collective nature, however the high-j quasiparticles  
from proton and neutron $(d_{5/2}g_{7/2})$ and $h_{11/2}$ orbitals   
do also substantially contribute by means of the  
shears mechanism. With increasing frequency, the 3-component 
of the spin vector $\vec J$ stays practically 
at $<J_3>\approx$\,9$\hbar$. However, this does not mean that the corresponding 
TAC configuration behaves like a structureless the high-K rotor. 
Fig.6 shows that the calculated dependence $\vartheta(\omega)$ of the tilt angle  
on the rotational frequency  by no means follows curve expected for
 the strong coupling limit.
Thus,  the present case lies in between a good shears band 
and a good high-K band.
 The $%
\Delta I=1$ band in $^{128}$Ba is an example for a rotational
band of  intermediate nature.

\section{Summary}
The strength of the TAC model is that it can predict 
the appearance of $\Delta I=1$ rotational bands and 
is able to describe microscopically their electro magnetic 
decay properties. This is achieved by taking into consideration
the  orientation of the rotational axis with respect to the deformed 
potential,
which is fixed  along a principal
axis 	in the conventional cranking model. The intrinsic TAC configuration
 of a rotational band
is found 
by searching for a local minimum on the  multi-parameter  surface
of the total Routhian. 
Therefore, it is crucial to calculate these surfaces
as reliable as possible. In the present work  the
Strutinsky renormalization and a hybrid single-particle potential
were implemented in order to improve
the total Routhian.
The hybrid potential combines the 
 spherical  Woods-Saxon single-particle
energies with the deformed part of the Nilsson potential. 
For moderately deformed nuclei it gives deformed single-particle levels
that are quite close to the deformed Woods-Saxon levels. 

We applied the hybrid TAC to the previously investigated  
 four quasiparticle magnetic dipole 
band in $^{128}$Ba.
 The lowest equilibrium configuration is
found to be  $[\pi( h_{11/2}(d_{5/2}g_{7/2}))\nu(
h_{11/2}(d_{5/2}g_{7/2}))]$.
It has positive parity and a prolate
axial deformation of $\varepsilon_2=0.20$.
The microscopic TAC calculations describes rather well the
experimental  energies,  $B(M1)$ and $B(E2)$ values, as well as the
branching and mixing ratios. The dipole band in $^{128}$Ba
has an intermediate structure. A
comparable amount of angular momentum is 
generated by the shears mechanism, active for the
$h_{11/2}$ and $(d_{5/2}g_{7/2})$ quasiparticles, and by collective rotation.
It is an example for the transition from collective to magnetic rotation.

 The hybrid potential turned out to be crucial 
for the good agreement between the calculation and the data. 
It substantially improves the Nilsson potential, which has problems in the 
region around mass 130.      
It seems to be a promising starting point for studying the intriguing
interplay between triaxial deformation and the orientation of the rotational
axis. 

\section{Acknowledgments}

V.D. wishes to thank the Foundation ''Bulgarian Science and Culture'' for
its support. 
We should like to thank Prof. P. von Brentano and Dr. I. Wiedenh\"over, who 
kept us up with the data.
 The work was partially carried out under the Grant DE-FG02-95ER40934.

\onecolumn

\begin{figure}[]
\mbox{\psfig{file=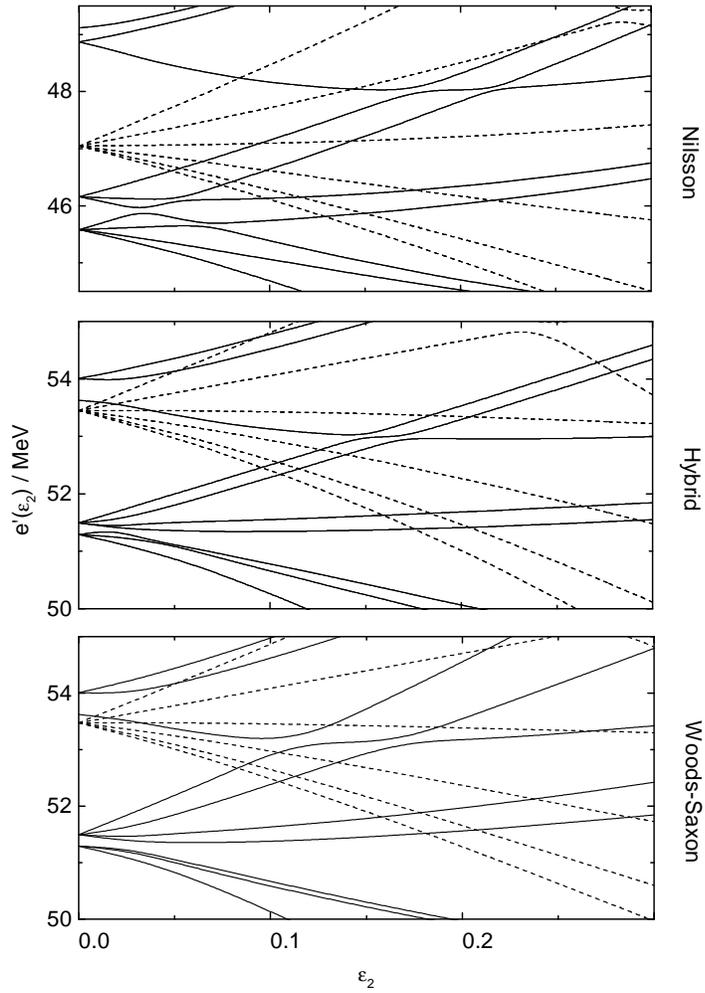,width=12cm}}
\vspace*{2cm}
\caption{ Proton single-particle levels as function of deformation for the
Woods-Saxon (bottom), hybrid (middle) and Nilsson (top) model, respectively. 
Full lines
present positive parity states, broken lines - negative parity states. The
Fermi energy lies at about 51.3 MeV for the Woods-Saxon and the hybrid
models and at 45.6 MeV for the Nilsson model. }
\end{figure}
\newpage
\begin{figure}[]
\mbox{\psfig{file=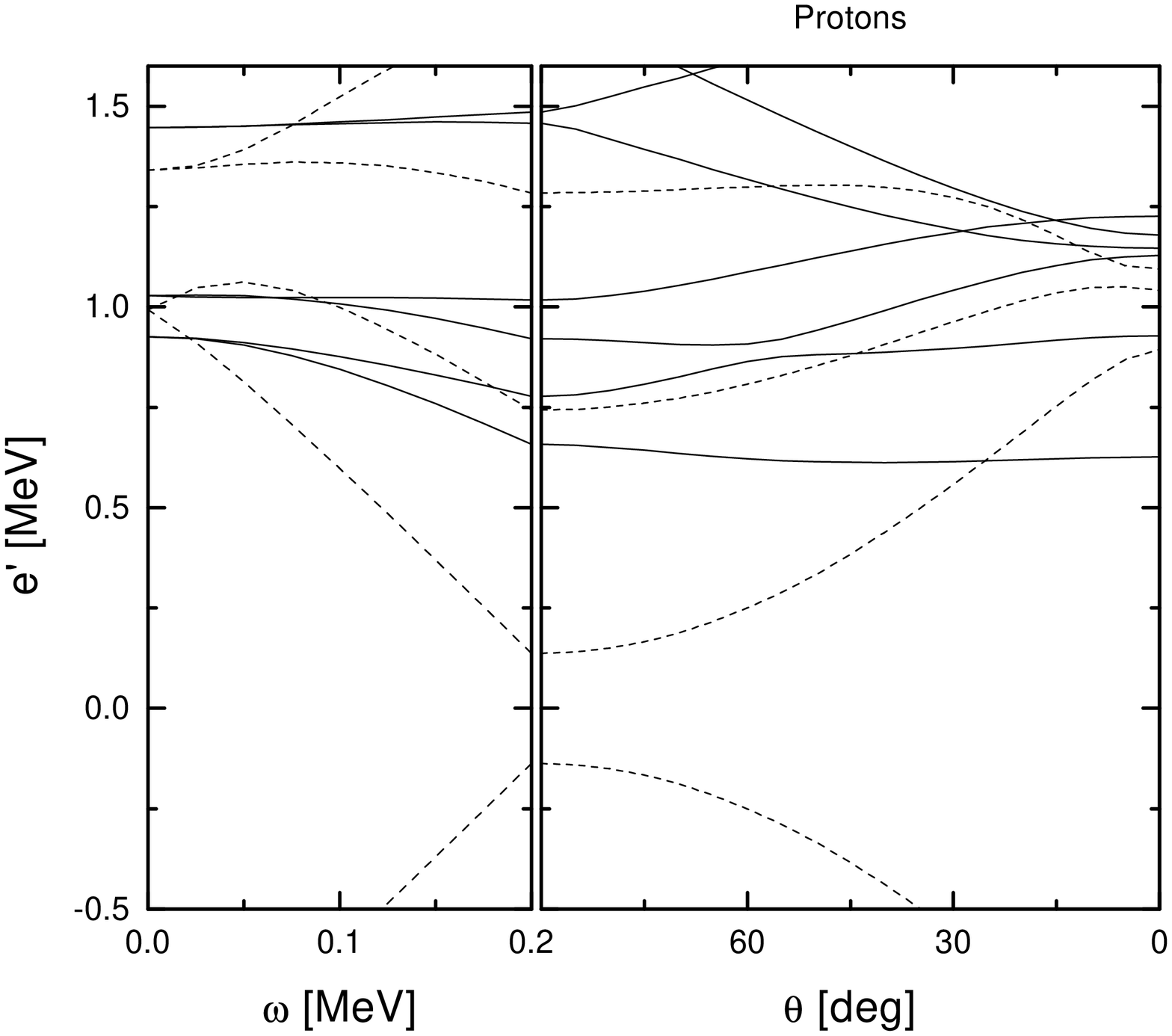,width=10cm}}
\mbox{\psfig{file=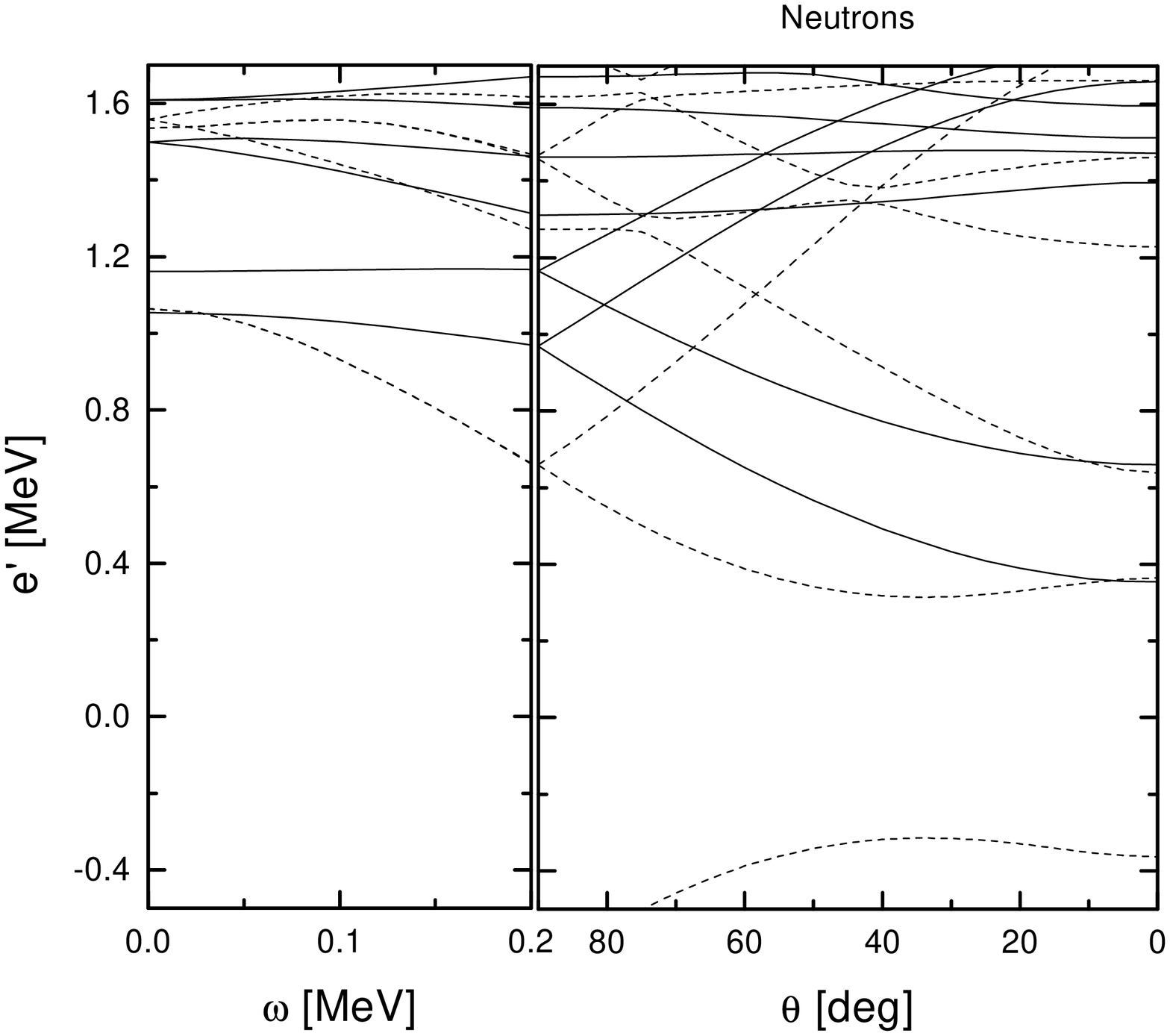,width=10cm}}
\vspace*{3cm}
\caption{ Proton (a) and neutron (b) quasi-particle levels as function of
rotational frequency
$\hbar\omega$ and tilt angle $\vartheta$. Full lines correspond to positive
parity, broken lines - to negative parity states. In both cases the lowest
two quasi-particle configurations involve one predominantly 
$(d_{5/2}g_{7/2})$ and
one predominantly $h_{11/2}$ state. The
 deformation parameters are 
$\varepsilon _{2}=0.205,\varepsilon_{4}=-0.01,\gamma =0^{o}$, which are the 
equilibrium values at $\omega=0.2~MeV$.}
\end{figure}
\newpage
\begin{figure}[]
\mbox{\psfig{file=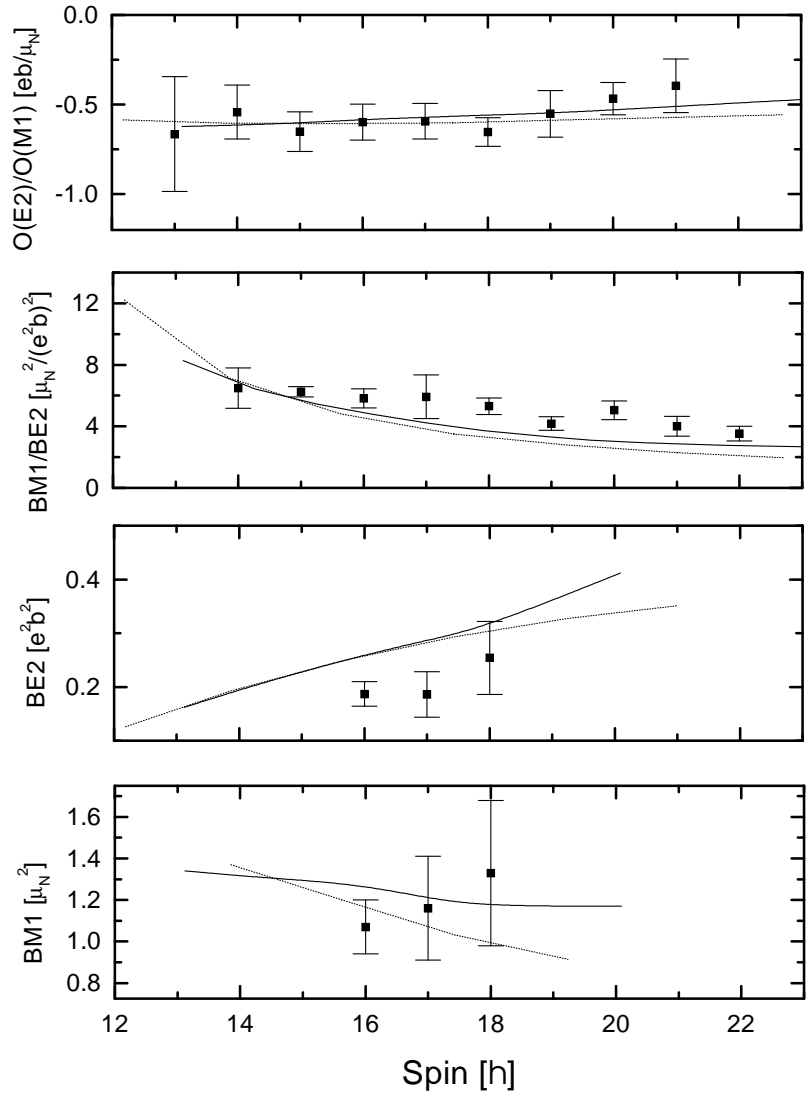,width=12cm}}
\vspace*{3cm}
\caption{ Electromagnetic observables as obtained by paired calculation
(full lines), unpaired calculation (dotted lines) and experiment.}
\end{figure}
\newpage
\begin{figure}[]
\mbox{\psfig{file=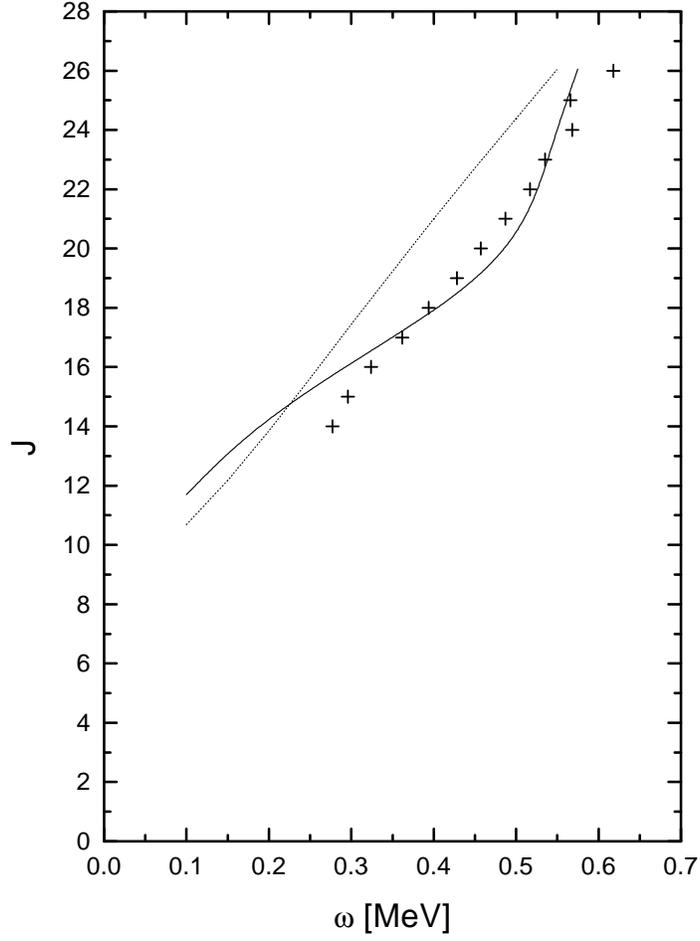,width=12cm}}
\vspace*{3cm}
\caption{ Dependence of the angular momentum $J=I$
 on the rotational frequency $\hbar\omega$
as obtained by
paired calculation (full line), unpaired calculation (dotted line) and
experiment (crosses). The experimental frequency is
extracted from the measured $\gamma-$energy data using the relations
$\hbar \omega(I) =E_{\gamma }=E(I)-E(I-1)$. 
}
\end{figure}
\newpage
\begin{figure}[]

\mbox{\psfig{file=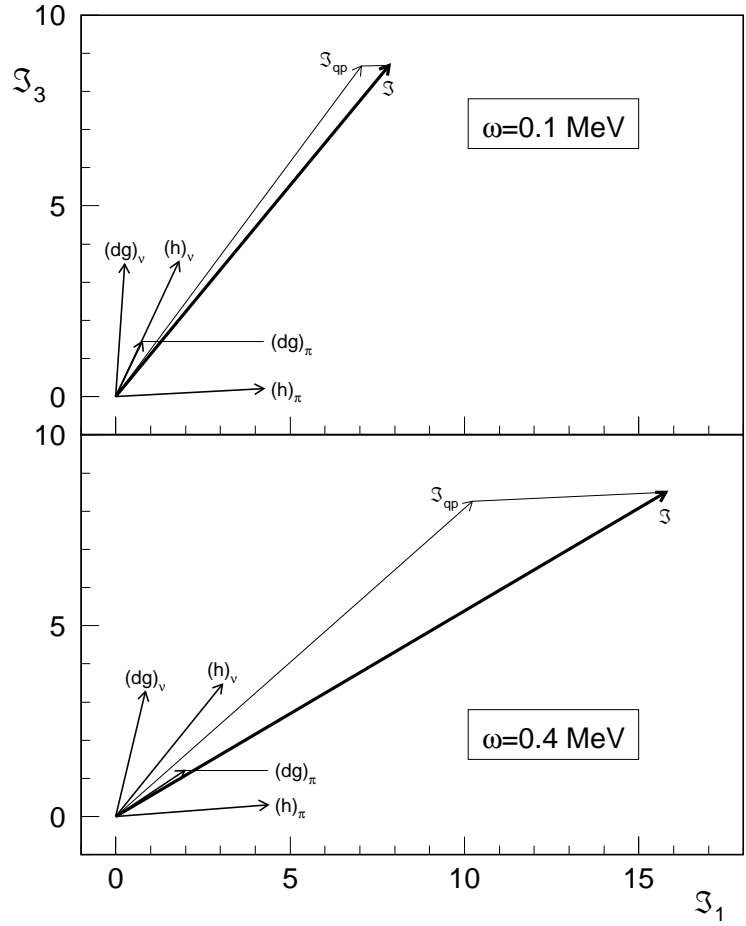,width=12cm}}
\vspace*{3cm}
\caption{ Spin decomposition in terms of individual quasi-particle
contributions for two different rotational frequencies. Quasi-particle
configurations are denoted by their predominant components. }
\end{figure}
\newpage
\begin{figure}[]
\mbox{\psfig{file=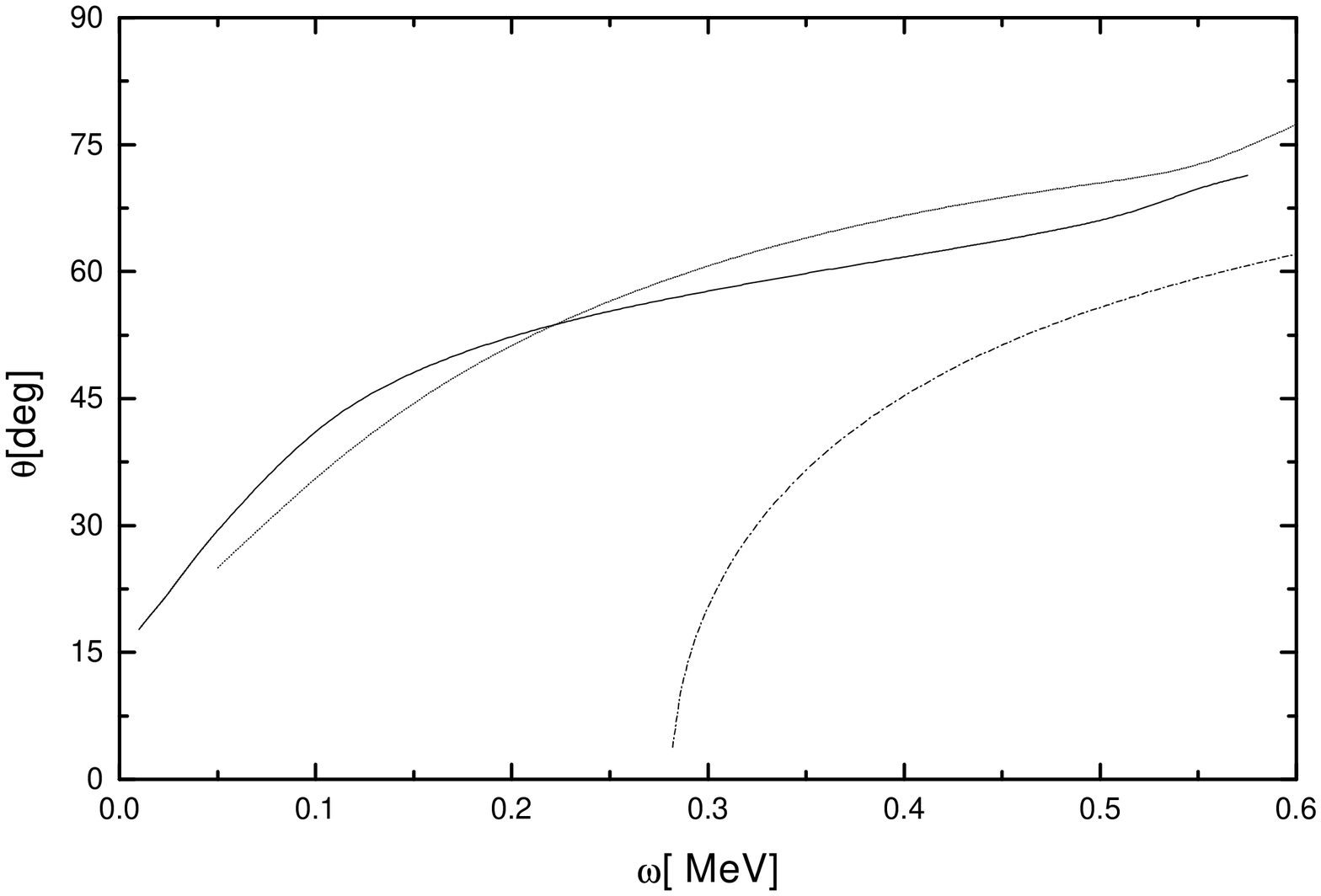,width=12cm}}
\vspace*{4cm}
\caption{ Equilibrium tilt angle $\vartheta$ as a function of the rotational frequency
 $\hbar\omega$. Full line -
paired calculation; dotted line - unpaired calculation; broken-dotted line -
strong coupling limit result. }
\end{figure}


\begin{references}
\bibitem{Ref1} see, for example D. Fossan, J. R. Hughes, Y. Liang,
R. Ma, E. S. Paul, and N. Xu,\\ {\it Nucl. Phys.} {\bf A520}, 241c  
\bibitem{Ref4}  S.Frauendorf, {\it Nucl.Phys}. {\bf A}557 (1993) p.259c
\bibitem{zphys} S. Frauendorf, Z. Phys. {\bf A385}, 163 (1997)
\bibitem{rmp} S. Frauendorf, Rev. Mod. Phys, submitted
\bibitem{tacdic} S. Frauendorf,{\it Nucl. Phys.} {\bf A}, submitted

\bibitem{Ref5}  P.Ring, P.Schuck; {\it The Nuclear Many Body Problem},
Springer NY 1980


\bibitem{Ref7}  F.D\"{o}nau, S.~Frauendorf, P.~von~Brentano, 
 A.~Gelberg, and O.~Vogel ,\\ {\it Nucl.Phys} {\bf A}584 (1995) p.241;
\bibitem{chiral} S. Frauendorf, and Meng, J., 1997,{\it Nucl. Phys. A} {\bf
617}, 131 (1997)
\bibitem{Ref7a} O.Vogel, A. Dewald, P. von Brentano, J. Gableske,
R. Kr\"ucken, N. Nicolay, A. Gelberg, P. Petkov,
A. Gizon, J. Gizon, D. Bazacco, C. Rossi Alvarez, S. Lunardi, P. Pavan,
D.R. Napoli, S. Frauendorf, and F. D\"onau , {\it Phys.Rev.} {\bf C}56 (1997) p.1338
\bibitem{Ref7b} P. Petkov, J. Gablenske, O. Vogel, A. Dewald, P. von Brentano, R. Kr\"ucken, R. Peusquens, N. Nicolay, A. Gizon, J. Gizon, D. Bazacco, C. Rossi-Alvarez, S. Lunardi, P. Pavan, D.R. Napoli, W. Andrejtscheff, R.V. Jolos,
{\it Nucl. Phys.} {\bf A} 640 (1998) 293 
\bibitem{Ref8} I. Wiedenh\"over, O. Vogel, H. Klein, A. Dewald, 
P. von Brentano, J. Gableske, R. Kr\"ucken, N. Nicolay,
A. Gelberg, P. Petkov, A. Gizon, D. Bazzacco, C. Rossi Alvarez,
G. de Angelis, S. Lunardi, P. Pavan, D.R. Napoli, S. Frauendorf, 
F. D\"onau, R.V.F. Janssens,and M. Carpender , {\it Phys.Rev}. {\bf C}58 (1998)
p.721
\bibitem{NilsRag} S.G. Nilsson and I. Ragnarsson, {\it Shapes and Shells in Nuclear Structure}, 
 Cambridge University Press, 1995

\bibitem{Ref9}  R. Bengtsson, private communication
\bibitem{Ref10} Wyss, R., Granderath, A., Bengtsson, R., von Brentano, P.,
Dewald, A., Gelberg, A., Gizon, A., Gizon, J., Harissopulos,
S., Johnson, A., Lieberz, W., Nazarewicz, W., Nyberg, J.,
and K. Schiffer, {\it Nucl. Phys. A} {\bf 505}, 337 (1989)
\bibitem{clhyb} S. Reimann, S. Frauendorf, M. Brack,{\it Z. Phys.} {\bf D 34},
125 (1995)
\bibitem{clks} B. Montag, Th. Hirschmann, J Meyer, P.-G. Reinhard, M. Brack,
\\
{\it Phys. Rev.} {\bf B 52}, 4775 (1995)

\bibitem{Ref13}  S.Cwiok, J. Dudek, W. Nazarevicz, J. Skalski, 
and T. Werner, {\it Comput. Phys.Commun.}. {\bf 46} (1987) 379

\bibitem{Ref14}  M.Ted Ressell, Maurice B. Aufderheide, Steward D. Bloom, 
Kim Griest, Grant J. Mathews, and David A. Resler {\it Phys.Rev}.{\bf \ D}48 (1993)
p.5519;\\ V.I.Dimitrov, J. Engel and S. Pittel, {\it Phys.Rev}.{\bf \ D}51 (1995) R291

\bibitem{BM} A. Bohr and B.M. Mottelson, {\it Nuclear Structure Vol. II},
W.A. Benjamin, Reading, MA, 1957
\end{references}
\end{document}